\documentclass[aps,10pt,prb,showpacs,twocolumn, superscriptaddress]{revtex4-1}
\usepackage{graphicx}
\usepackage{amssymb}
\usepackage{amsfonts}
\usepackage{color}
\usepackage{bm}
\newcommand\varpm{\mathbin{\vcenter{\hbox{%
  \oalign{\hfil$\scriptstyle+$\hfil\cr
          \noalign{\kern-.3ex}
          $\scriptscriptstyle({-})$\cr}%
}}}}
\newcommand\varmp{\mathbin{\vcenter{\hbox{%
  \oalign{$\scriptstyle({+})$\cr
          \noalign{\kern-.3ex}
          \hfil$\scriptscriptstyle-$\hfil\cr}%
}}}}
\DeclareMathAlphabet      {\mathbf}{OT1}{cmr}{bx}{n}

\begin{document}

\title{Ghost modes and continuum scattering in the dimerized distorted kagome lattice antiferromagnet Rb$_2$Cu$_3$SnF$_{12}$}
\author{K.~Matan}
\email{kittiwit.mat@mahidol.ac.th}
\affiliation{Department~of~Physics,~Faculty~of~Science,~Mahidol~University, Bangkok 10400, Thailand}
\affiliation{ThEP,~Commission of Higher Education,~Bangkok 10400, Thailand}
\author{Y.~Nambu} 
\affiliation{IMRAM, Tohoku University, Sendai, Miyagi 980-8577, Japan}
\author{Y.~Zhao}
\affiliation{Department of Materials Science and Engineering, University of Maryland, College Park, MD 20742, USA}
\affiliation{NIST Center for Neutron Research, National Institute of Standards and Technology, Gaithersburg, MD 20899, USA}
\author{T.~J.~Sato}
\affiliation{IMRAM, Tohoku University, Sendai, Miyagi 980-8577, Japan}
\author{Y.~Fukumoto}
\affiliation{Department of Physics, Faculty of Science and Technology, Tokyo University of Science, Noda, Chiba 278-8510, Japan}
\author{T.~Ono}
\affiliation{Department of Physical Science, School of Science, Osaka Prefecture University, Sakai, Osaka 599-8531, Japan}
\author{H.~Tanaka}
\email{tanaka@lee.phys.titech.ac.jp}
\affiliation{Department of Physics, Tokyo Institute of Technology, Meguro-ku, Tokyo 152-8551, Japan}
\author{C.~Broholm}
\affiliation{NIST Center for Neutron Research, National Institute of Standards and Technology, Gaithersburg, MD 20899, USA}
\affiliation{Institute for Quantum Matter and Department of Physics and Astronomy, The Johns Hopkins University, Baltimore, MD 21218, USA}
\affiliation{Quantum Condensed Matter Division, Oak Ridge National Laboratory, Oak Ridge, Tennessee 37831, USA}
\author{A.~Podlesnyak}
\author{G.~Ehlers}
\affiliation{Quantum Condensed Matter Division, Oak Ridge National Laboratory, Oak Ridge, Tennessee 37831, USA}

\date{\today}
\begin{abstract}
High intensity pulsed neutron scattering reveals a new set of magnetic excitations in the pinwheel valence bond solid state of the distorted kagome lattice antiferromagnet Rb$_2$Cu$_3$SnF$_{12}$.  The polarization of the dominant dispersive modes (2~meV$<\hbar\omega<$7~meV) is determined and found consistent with a dimer series expansion with strong Dzyaloshinskii-Moriya interactions ($D/J=0.18$).  A weakly dispersive mode near 5~meV and shifted ``ghosts'' of the main modes are attributed to the enlarged unit cell below a $T=215$ K structural transition.  Continuum scattering between 8~meV and 10~meV might be interpreted as a remnant of the kagome spinon continuum [T.-H. Han \textit{et al.}, Nature {\bf 492}, 406 (2012)].
\end{abstract}
\maketitle

\section{Introduction}
Interacting spins on the two-dimensional kagome lattice have fascinated physicists since Syozi first showed that Ising spins on this lattice, which he named after the woven pattern on a Japanese bamboo basket, do not order for $T\rightarrow0$~(ref.~\onlinecite{Syozi:1951tm}).  More recently, efforts have focused on determining the ground state of the quantum \mbox{spin-$\frac{1}{2}$} Heisenberg kagome lattice antiferromagnet, which is considered to be one of the most challenging problems in condensed matter physics.  The complexity arises from the macroscopic degeneracy caused by the incompatibility between the global geometry of the corner-sharing triangular network and local, nearest-neighbor antiferromagnetic interactions.\cite{ramirez}  The classical N\'{e}el state is apparently replaced by a dynamic quantum state, the details of which remain to be established. Proposed ground states include a gapless $U(1)$-Dirac-spin-liquid state,\cite{Hastings:2001p247,Ran:2007p28,Iqbal:2013cl,Iqbal:2012fd} a gapped-spin-liquid,\cite{Sachdev:1992p773,Waldtmann:1998p817,ZENG:1995p1054,Misguich:2002p3766,Depenbrock:2012ho} and valence-bond-solid (VBS) states.\cite{MARSTON:1991p327,Nikolic:2003p879,Singh:2007p3664,Singh:2008cn,Yang:2008p462} These states are very close in energy so small perturbations and intrinsic limitations of numerical methods make it difficult to reach a firm conclusion.  Most of the recent theoretical studies point to a quantum spin liquid,\cite{Yan:2011kt} although there is no consensus on its precise nature.

Identifying an ideal kagome lattice model system has also proven to be difficult.  All realizations so far have been plagued by magnetic impurities, lattice distortion, and extra terms in the spin Hamiltonian including anisotropic and further-neighbor interactions.\cite{Helton:2007p2715,Helton:2010ci,Lee:2007p1219,Han:2012cz,Han:2011ch,Ofer:2009p3188,Mendels:2007p335,Bert:2005p1111,Yoshida:2009p5506,Nilsen:2011ep,Fak:2012iu}  Albeit minuscule in some cases, these effects may conceal the intrinsic nature of the nearest-neighbor Heisenberg kagome antiferromagnet (HKAFM).  Still, much can be learned by studying materials with interacting quantum spins on kagome-like lattices.  For the quantum spin-$\frac{1}{2}$ kagome lattice antiferromagnet ZnCu$_3$(OH)$_6$Cl$_2$ (herbertsmithite), a recent experiment by Han et al. indicates that fractionalized excitations, a key characteristic of spin liquids, are robust against a small excess of Cu$^{2+}$ ions in the interlayer sites and against anisotropic Dzyaloshinskii-Moriya (DM) interactions.\cite{han_nature}  The recent discovery of the pinwheel VBS state in the distorted kagome lattice antiferromagnet Rb$_2$Cu$_3$SnF$_{12}$ offers a rare opportunity to study a cooperative singlet on an  approximate kagome lattice.\cite{matan_natphy}  Besides being unique and interesting in its own right, the pinwheel VBS state may display intersite correlations and excitations related to the ideal HKAFM.\cite{khatami}  

\begin{figure}
\centering \vspace{0in}
\includegraphics[width=8cm]{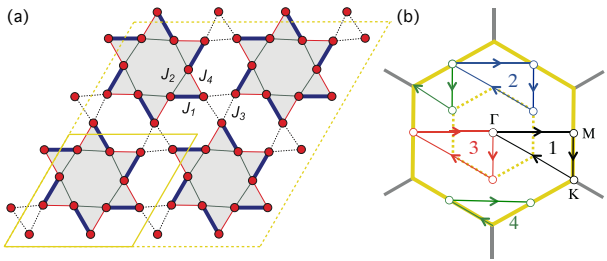}
\caption{(Color online) (a) The pinwheel VBS state is formed by dimers (thick lines). A dimer is a pair of spins with the largest exchange interaction $J_1$.  The exchange interactions are $J_1>J_2>J_3>J_4$. Yellow solid lines denote a two-dimensional unit cell of the room-temperature phase while yellow dotted lines denote the $2a\times2a$ enlarged unit cell.  (b) A diagram showing paths 1, 2, 3, and 4 in the first Brillouin zone.  The dotted hexagon denotes the smaller Brillouin zone associated with the enlarged unit cell.}\label{fig1}
\end{figure}

At room temperature, Rb$_2$Cu$_3$SnF$_{12}$ has the hexagonal $R\bar{3}$ space group with lattice parameters $a=13.917(2)~$\AA~ and $c=20.356(3)~$\AA.\cite{ono:174407} At 215 K, it undergoes a first-order structural transition, doubling the in-plane lattice constant $a$. The resulting lattice distortion is small\cite{matan_natphy} so to a first approximation we use the room temperature structure, where a two-dimensional unit cell comprises 12 Cu$^{2+}$ spins (Fig.~\ref{fig1}(a)). The spin-$\frac{1}{2}$ Cu$^{2+}$ ions form a distorted kagome plane and are surrounded by a deformed octahedral environment of fluorine.  The kagome planes are separated by nonmagnetic ions, which results in weak interlayer interactions.  The distorted kagome lattice gives rise to four antiferromagnetic in-plane exchange interactions $J_{1}>J_{2}>J_{3}>J_{4}$ (Fig.~\ref{fig1}(a)). To lowest order, spins interacting through $J_1$ form singlets which are linked through the weaker interactions.  Powder neutron diffraction shows no magnetic order down to 1.3 K.  The low-temperature magnetic susceptibility indicates a non-magnetic, spin singlet ($S_{\mathrm{tot}}=0$) ground state and mixing of the singlet and triplet ($S_{\mathrm{tot}}=1$) excited states through the DM interactions\cite{Morita:2008p2717} ($S_{\mathrm{tot}}$ denotes the quantum number for the total spin of a single dimer). To a good approximation, the spin Hamiltonian is given by
\begin{equation}
{\cal H} = \sum_{nn} \left[ J_{ij} {\bf S}_i \cdot {\bf S}_j + {\bf D}_{ij}\cdot {\bf S}_i \times {\bf S}_j \right],\label{eq1}
\end{equation}
where $J_{ij}>0$ are the nearest-neighbor antiferromagnetic exchange interactions, and ${\bf D}_{ij}$ are the corresponding DM vectors. 

In a previous study,\cite{matan_natphy} involving several of the present authors, magnetic excitations from the singlet ground state were probed using inelastic neutron scattering on a triple-axis spectrometer. These measurements revealed the pinwheel motif of dimers, and determined the relevant spin Hamiltonian parameters through a dimer series expansion up to eighth order.  However, the detailed structure of the excitations could not be resolved due to lack of resolution and counting statistics (see Fig.~2 in ref.~\onlinecite{matan_natphy}).  Here we report high intensity pulsed neutron scattering measurements on single crystalline Rb$_2$Cu$_3$SnF$_{12}$ using the Cold Neutron Chopper Spectrometer at the Spallation Neutron Source, Oak Ridge National Laboratory.\cite{Ehlers:2011ex} A time-resolved, highly pixelated detector system that covers a large solid angle (14\% of the unit sphere) enabled concurrent measurements over a much wider range of momentum and at higher resolution than previously. We confirm the splitting of the triplet associated with dimerization into a doublet and a singlet as a result of strongly anisotropic interactions, and are able to unambiguously determine the polarization of each mode.  More importantly, we discover a new family of modes associated with the structural superlattice, and a continuum at high energy, which may be related to the spinon continuum recently detected in the undimerized kagome system, herbertsmithite.\cite{han_nature}  

The article is organized as follows: In Sec.~\ref{sec2}, we describe the inelastic neutron scattering experiment and the resulting data.  In Sec.~\ref{sec3A}, the measured magnetic excitations are analyzed in the framework of a dimer series expansion for the $2a\times2a$ enlarged unit cell.  We find very good agreement for energy transfer less than 8~meV.  We analyze the wave vector dependence of scattering perpendicular to the kagome planes to determine the magnetic polarization of each mode in Sec.~\ref{sec3B}.  This confirms that the triplet is split into a singlet and a doublet.  Sec.~\ref{sec3C} is devoted to a discussion of the excitation spectrum and continuum scattering between 8~meV and 10~meV, which cannot be explained by the dimer series expansion.  We end with a summary in Sec.~\ref{sec4}

\begin{figure}
\centering \vspace{0in}
\includegraphics[width=8cm]{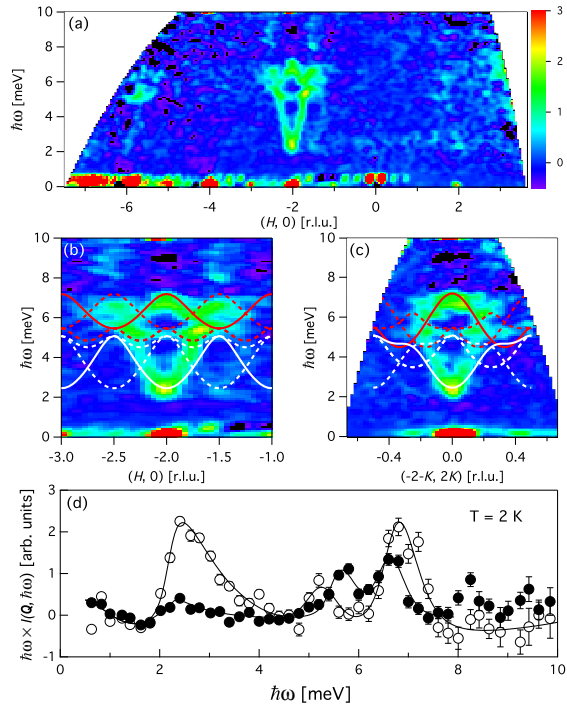}
\caption{(Color online) Contour maps show a product of scattering intensity and energy transfer $\hbar\omega\times I({\bf Q},\hbar\omega)$, displaying magnetic excitations in Rb$_2$Cu$_3$SnF$_{12}$ as a function of energy and in-plane momentum through (a), (b)~$(H, 0)$ and (c)~$(-2-K, 2K)$.  The intensity is averaged over the available range of $L$, $\Delta{\bf Q}_{[1,0]}$ of 0.042~\AA$^{-1}$, and $\Delta{\bf Q}_{[-1,2]}$ of 0.045~\AA$^{-1}$. The measurements along $[-K,2K]$ are limited by a smaller detector-coverage area perpendicular to the horizontal plane. Solid lines represent the excitations of the original spin Hamiltonian whereas dotted lines denote excitations resulting from the $2a\times2a$ enlarged unit cell.  Red denotes the $S_{\mathrm{tot},z}=0$ mode and white denotes the $S_{\mathrm{tot},z}=\pm1$ mode. (d) Constant-${\bf Q}$ cuts show $\hbar\omega\cdot I({\bf Q},\hbar\omega)$ at $(-2,0)$ (open circles) and the average of $\hbar\omega\cdot I({\bf Q},\hbar\omega)$ at $(-1.5,0)$ and $(-2.5,0)$ (closed circles).  Above 8~meV, the closed symbols lie above background, indicative of the continuum scattering.  The lines are guides to the eye.  The error bar represents one standard deviation.}\label{fig2}
\end{figure}

\section{Experimental and results}\label{sec2}

Single crystalline Rb$_2$Cu$_3$SnF$_{12}$ was synthesized from the melt using the method described in ref.~\onlinecite{Morita:2008p2717}. Inelastic neutron scattering measurements were performed on two co-aligned crystals with a total mass of 4~g and a mosaic of 1.5$^\circ$. The sample was mounted with the $(H,0,L)$ reciprocal lattice plane horizontal to allow intensity integration of rod-like scattering along the $L$ direction while taking advantage of the two dimensionality of the system.  The incident energy $E_i$ was fixed at 12~meV for an energy resolution (full width at half maximum) of 0.56(3)~meV at the elastic position.  The sample was cooled to a base temperature of 2 K using a He-4 cryostat.  Multiple datasets were acquired by rotating the sample about the vertical axis, which is parallel to $[-1,2,0]$, in steps of 2$^\circ$ covering 68$^\circ$ of sample orientation.  An angle between the incident beam and $[0,0,1]$ ranges from $-28.5^\circ$ to $39.5^\circ$. The background was measured at 70 K, where the excitations are very broad and weak.\cite{matan_natphy}  These datasets were subsequently combined to produce a background-subtracted, four-dimensional scattering-intensity function $I({\bf Q},\hbar\omega)$, where ${\bf Q}$ is the momentum transfer and $\hbar\omega$ is the energy transfer.  The data were sliced and cut along high-symmetry directions using MSLICE\cite{dave} to produce contour maps,  constant-${\bf Q}$, and constant-energy plots.

A contour map of $\hbar\omega\times I({\bf Q},\hbar\omega)$ averaged over the $L$ direction (the $L$-dependence of the scattering intensity will be discussed later), which is plotted as a function of energy and in-plane momentum along $[H,0]$ (Fig.\ref{fig2}(a)), shows a distinct pattern of excitations around $(-2,0)$ and faint outlines of similar patterns displaced by $\Delta H=\pm4$.  The latter are barely detectable around the equivalent Brillouin-zone centers, $(-6,0)$ and $(2,0)$. The measurements were set up so integration along $L$ is optimal at $(-2,0)$.  The difference in the intensity profile around $(-2,0)$ and $(2,0)$ is a result of a smaller range of intensity integration for the latter. The overall profile of the excitations around $(-2,0)$ is consistent with our previous report.\cite{matan_natphy} The lower branch, which has a broader band-width, is known to be a two-fold-degenerate excitation as it is split by a magnetic field along the $c$-direction.\cite{matan_natphy}  By mapping the $L$-dependence of the intensity of this branch, we shall later show that it is associated with transitions from the singlet ground state ($S_{\mathrm{tot}}=0$) to a doublet with $S_{\mathrm{tot}}=1$ and $S_{\mathrm{tot},z}=\pm1$ ($S_{\mathrm{tot},z}$ denotes the magnetic quantum number of the $S_{\mathrm{tot}}=1$ triplet states). The upper branch, which has a smaller band-width, does not split in a field and so is thought to be a non-degenerate excitation from the singlet ground state to the singlet state with $S_{\mathrm{tot}}=1$ and $S_{\mathrm{tot},z}=0$. We note that these states  are not pure states of defined angular momentum due to the DM interactions.  

A constant-${\bf Q}$ cut at $(-2,0)$ (Fig.~\ref{fig2}(d)) shows clear resolution-limited peaks at $\hbar\omega=2.4(3)$~meV and 6.9(3)~meV, consistent with the previous data (where the uncertainty represents half the energy resolution). The contour maps around the zone center $(-2,0)$ (Fig.~\ref{fig2}(b),~(c)) reveal a more intricate set of excitations than previously appreciated.  A weakly dispersive mode around 5~meV is visible along both $[H,0]$ and $[-K,2K]$ (Fig.~\ref{fig2}(b),~(c)).  At the zone center this mode peaks at 5.3(3)~meV (Fig.~\ref{fig2}(d)). It grows slightly more intense away from the zone center, which contrasts with the other two modes that become weaker. We also observe excitations centered around $(-1.5,0)$ and $(-2.5,0)$ (Figs.~\ref{fig2}(b) and~\ref{fig5}(a) ($2.0-2.5$~meV)), which resemble the mode around $(-2,0)$ but with much less intensity, and hence are named the `ghost' modes.  We have previously reported these ghost modes and attributed them to the enlarged unit cell caused by the structural transition.\cite{matan_natphy}  Our dimer series expansion shown by solid lines in Fig.~\ref{fig2}(b),~(c) and the bond-operator mean-field theory\cite{Hwang:2012iv} cannot account for all of this observed scattering intensity between 2~meV and 7~meV as neither calculation considers the enlarged unit cell.  Furthermore, we observe diffuse scattering between 8~meV and 10~meV near $(-1.5,0)$ and $(-2.5,0)$, which cannot be accounted for by the dimer series expansion.

\section{Analysis and Discussion}
In this section, we first analyze the excitation dispersions below 8~meV using the extended version of the dimer series expansion, which has been discussed in our previous work\cite{matan_natphy}, to include the effect of the enlarge unit cell.  We then investigate the $L$-dependence of the intensity to determine the polarization of each mode.  We end this section with a discussion of the diffuse scattering between 8~meV and 10~meV.

\begin{figure}
\centering \vspace{0in}
\includegraphics[width=8cm]{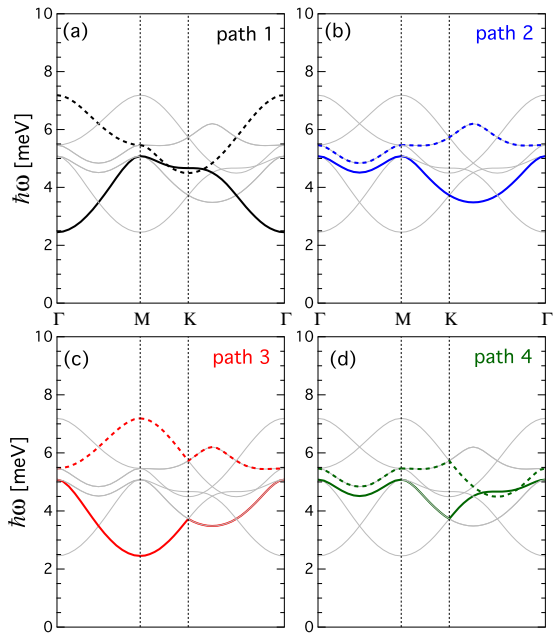}
\caption{(Color online) (a) Singlet-to-triplet excitations along path 1 (Fig.~\ref{fig1}(b)) for the unperturbed Hamiltonian ${\cal H}$ are shown by black lines.  Gray lines denote the modes resulting from the perturbed Hamiltonian ${\cal H'}$ associated with the enlarged unit cell.  Solid lines denote excitations from the singlet ground state to the $S_{\mathrm{tot},z}=\pm1$ states whereas dotted lines denote excitations to the the $S_{\mathrm{tot},z}=0$ state. Colored  solid and dotted lines in (b), (c), and (d) show the excitations along paths 2, 3, and 4 (Fig.~\ref{fig1}(b)), respectively.}\label{fig3}
\end{figure}

\subsection{Dimer series expansion and enlarged unit cell}\label{sec3A} 

To understand the 5~meV mode and the ghost modes around $(-1.5,0)$ and $(-2.5,0)$, we consider the $2a\times2a$ enlarged unit cell consisting of 48 spins shown in Fig.~\ref{fig1}(a).  We write the spin Hamiltonian as ${\cal H + H'}$, where ${\cal H'}$ represents a perturbation due to the enlarged unit cell.  $\cal H'$ has the exact same form as $\cal H$ (eq.~\ref{eq1}) but the sum is over 48 spins in the enlarged unit cell (see Fig.~\ref{fig1}(a)).  We then perform the dimer series expansion on the pinwheel VBS state using the Hamiltonian ${\cal H + H'}$.  The linked cluster expansion algorithm was used to generate a graphical series of dimers.\cite{oitmaa}  The low-energy spectra are calculated up to eighth order in the inter-dimer and DM interactions using the Dlog-Pad\'e approximation.\cite{matan_natphy}  We define the path $\Gamma\rightarrow M\rightarrow K\rightarrow \Gamma$ in the first Brillouin zone of the original model ${\cal H}$ as path 1 (Fig.~\ref{fig1}(b)).  The lowest energy excitations with $S_{\mathrm{tot},z}=\pm1$ and those with $S_{\mathrm{tot},z} = 0$ along path 1 are shown in Fig.~\ref{fig3}(a).  We also define paths 2, 3, and 4 (Fig.~\ref{fig1}(b)), which differ by a reciprocal lattice vector of the enlarged unit cell. Dispersion curves for paths 2, 3, and 4 are shown in Fig.~\ref{fig3}(b), (c), and (d), respectively.  In Fig.~\ref{fig2}(b), (c) path 1 shown as solid lines and paths 2, 3, and 4 shown as dotted lines are qualitatively in agreement with the data. If ${\cal H'}$ is non-zero but very small, then paths 2, 3, and 4 become equivalent to path 1, and the dispersion curves shown by dotted lines are the anticipated ghost modes together with the corresponding excitations of the original Hamiltonian.  

It is interesting to note that the weakly dispersive mode in Fig.~\ref{fig2}(b) is originally the excitation on the path between two adjacent $M$ points (paths 2 and 4 in Fig.~\ref{fig1}(b)). From the experiment, this mode is not symmetric around $H = -1.75$; its energy increases monotonically as $H$ varies from $-2$ to $-1.5$. This suggests that we observe the $S_{\mathrm{tot},z} = \pm1$ triplet excitations around the zone center, $H = -2$, and the $S_{\mathrm{tot},z} = 0$ triplet excitations away from there, which may be experimentally verified by measurements in a magnetic field. We note that the in-plane component of the DM vector $d_p$ is set to zero in our dimer series expansion. A recent $^{63,65}$Cu NMR study in high fields up to 30 T\cite{grbic} and neutron scattering measurements\cite{nambu} show that the mixing between the singlet and triplet states via the DM interactions gives rise to a large residual gap. The anti-crossing of the singlet and triplet mode, which is due to the combined effect of the off-diagonal $g$-tensor and small $d_p$ ($|d_p|<0.012$),\cite{grbic} prevents the gap from closing at high magnetic fields.  However, $d_p$ has little effect on the overall zero-field spectrum.\cite{matan_natphy} 

\begin{figure*}
\centering \vspace{0in}
\includegraphics[width=12cm]{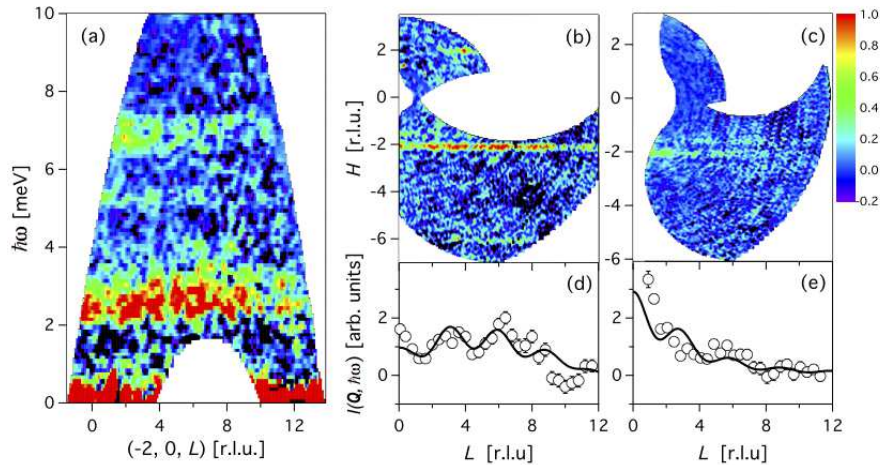}
\caption{(Color online) (a) Contour maps of $I({\bf Q},\hbar\omega)$ are plotted as a function of $\hbar\omega$ and $L$. The intensity is averaged over $\Delta{\bf Q}_{[1,0]}$ of 0.042~\AA$^{-1}$ and $\Delta{\bf Q}_{[-1,2]}$ of 0.045~\AA$^{-1}$.  The broadening is a result of the integration along the in-plane momenta.  (b) and (c) show contour maps of $I({\bf Q},\hbar\omega)$ plotted as a function of $H$ and $L$. The intensity is energy-integrated for (b) $\hbar\omega=[2.0,3.0]$~meV and (c) $\hbar\omega=[6.5,7.5]$~meV.  (d) and (e) show the $L$-dependence of $I({\bf Q},\hbar\omega)$ for energy ranges (d) $\hbar\omega=[1.0,4.0]$~meV and (e) $\hbar\omega=[6.0,8.0]$~meV centered at $(-2,0)$. The intensity is averaged over $\Delta{\bf Q}_{[1,0]}$ of 0.10~\AA$^{-1}$ and $\Delta{\bf Q}_{[-1,2]}$ of 0.18~\AA$^{-1}$. Solid lines denote the product of the magnetic form factor for Cu$^{2+}$ spins, the inter-plane correlation function, and  the polarization factor, assuming that modes are polarized (d) in the kagome plane and (e) out of the plane.}\label{fig4}
\end{figure*}

\subsection{$L$-dependence and mode polarization}\label{sec3B}
The scattering intensity displayed thus far was averaged over the $L$ direction.  However, the $L$-dependence of the scattering intensity contains valuable information about the polarization of the excitations and inter-plane correlations. Within the resolution of our measurements, there is no dispersion along $L$ (Fig.~\ref{fig4}(a)), which attests to the two-dimensional nature of the system. Contour maps of the scattering intensity integrated over the energy ranges $\hbar\omega=[2.0,3.0]$~meV (Fig.~\ref{fig4}(b)) and $\hbar\omega=[6.5,7.5]$~meV (Fig.~\ref{fig4}(c)) plotted as a function of $H$ and $L$ show rods of scattering that extend along $L$ at $H=-6$, $-2$, and 2. The integrated intensity of the $S_{\mathrm{tot},z}=\pm1$ mode has broad maxima at $L=3$ and $6$ before falling off at large $L$ (Fig.~\ref{fig4}(d)), while that of the $S_{\mathrm{tot},z}=0$ mode monotonically decreases as a function of $L$ with a small hump around $L=6$ (Fig.~\ref{fig4}(e)). The overall trend of the curves reflects the different polarization of the modes while the modulation of scattering intensity results from inter-plane correlations. The magnetic scattering cross section\cite{lovesey} can be described by
\begin{widetext}
\begin{equation}
\frac{d^2\sigma}{d\Omega~dE'}=N_M\frac{\mathbf{k'}}{\mathbf{k}}\left(\gamma r_0\right)^2\left[\frac{g}{2}f(\mathbf{Q})e^{-W}\right]^2\sum_{\alpha,\beta}\left(\delta_{\alpha\beta}-\hat{Q}_\alpha\hat{Q}_\beta\right)S^{\alpha\beta}\left(\mathbf{Q},\omega\right),
\end{equation}
\end{widetext}
where the $\mathbf{Q}$-dependent terms are the magnetic form factor $f(\mathbf{Q})$ and the dynamic magnetic structure factor $S^{\alpha\beta}\left(\mathbf{Q},\omega\right)$, which is the space and time Fourier transform of the spin-pair correlation function. The magnetic scattering cross section also contains the polarization factor that arises from the anisotropy of the dipole-dipole interaction between neutrons and electrons.  

For the magnetic excitations in Rb$_2$Cu$_3$SnF$_{12}$, the polarization factor becomes $1\varpm(Q_{L}/|{\bf Q}|)^2$, where $Q_{L}$ is a component of ${\bf Q}$ along $L$.  It grows (shrinks) with increasing $L$ if the polarization is in-plane or transverse (out-of-plane or longitudinal).\cite{lovesey}  The inter-plane correlations, which are embedded in the dynamic structure factor, can be described by a function $1+\alpha\cos\left(\frac{2\pi L}{3}\right)$, when the correlations along $c$ only extend to the nearest neighbor plane located at $\frac{c}{3}$.  Here the fit parameter, $\alpha$, indicates the type and strength of inter-plane correlations. (Ferromagnetic for positive $\alpha$.  Antiferromagnetic for negative $\alpha$.)   The product of the magnetic form factor for Cu$^{2+}$ spins, which decreases monotonically with increasing $L$, the polarization factor, and the inter-plane correlation function denoted by a solid line in Fig.~\ref{fig4}(d) (Fig.~\ref{fig4}(e)) is in accordance with the in-plane (out-of-plane) polarization of the $S_{\mathrm{tot},z}=\pm1$ ($S_{\mathrm{tot},z}=0$) mode.  Ferromagnetic inter-plane correlations are indicated by positive $\alpha$ ($\alpha=0.31(15)$).  The polarization analysis for the excitations around 5~meV close to the zone boundary, which is not shown, reveals mixing of the in-plane and out-of plane polarizations, or in other words the $S_{\mathrm{tot},z}=0$ and $S_{\mathrm{tot},z}=\pm1$ modes merge near the zone boundary.

\begin{figure}
\centering \vspace{0in}
\includegraphics[width=8cm]{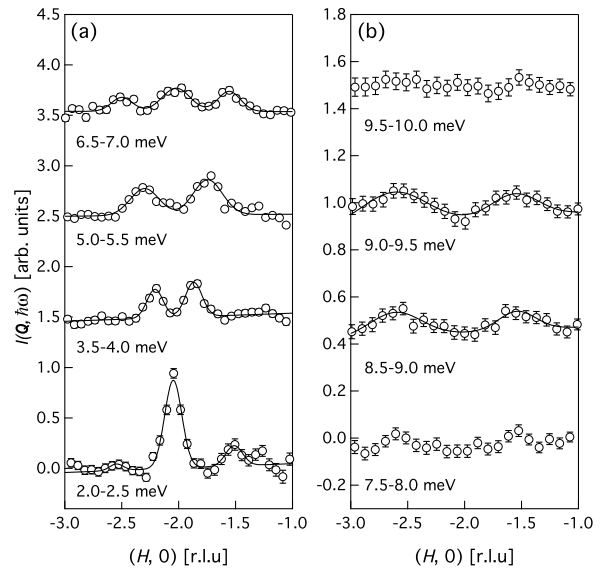}
\caption{Constant-energy cuts of $I({\bf Q},\hbar\omega)$ are plotted as a function of $H$. The energy range of the integration is indicated below each data set. The intensity is averaged over the whole range of $L$ and $\Delta{\bf Q}_{[-1,2]}$ of 0.045~\AA$^{-1}$. The lines serve as guides to the eye. Data sets for different energy ranges in (a) and (b) are shifted vertically by 1 (1.5 for 2.0-2.5~meV) and 0.5, respectively.}\label{fig5}
\end{figure}

\subsection{Continuum scattering}\label{sec3C}

The magnetic excitation spectrum between 2~meV and 7~meV in Rb$_2$Cu$_3$SnF$_{12}$ is markedly different from the spin-wave excitations observed in the classical spin-$\frac{5}{2}$ kagome lattice antiferromagnet KFe$_3$(OH)$_6$(SO$_4$)$_2$ (jarosite), which orders magnetically at low temperatures.\cite{Matan:2006p2713} It also differs from the continuum of spinon excitations in herbertsmithite, where the ground state is believed to be a quantum spin liquid.\cite{han_nature}  Resonant modes in Rb$_2$Cu$_3$SnF$_{12}$ (Fig.~\ref{fig5}(a)) are found only around the zone center and their intensity decreases precipitously away from $(-2,0)$, while in jarosite the `weathervane' mode exists throughout the Brillouin zone and in herbertsmithite the spinon continuum gives rise to hexagonal rings of diffuse scattering, surrounding zone centers.\cite{han_nature} However, between 8~meV and 10~meV we observe for Rb$_2$Cu$_3$SnF$_{12}$ weak diffuse scattering near $(-1.5, 0)$ and $(-2.5, 0)$ (Figs.~\ref{fig2}(b), (d) and ~\ref{fig5}(b)) on both sides of the zone center $(-2,0)$, as in herbertsmithite.  This scattering, which is diffuse in energy and broad in momentum, is different from the resolution-limited excitations below 8~meV and cannot be accounted for within the dimer series expansion.  For herbertsmithite, the recent neutron scattering places an upper bound of 0.25~meV on any gap in the continuum of scattering.\cite{han_nature}  On the contrary, the continuum in Rb$_2$Cu$_3$SnF$_{12}$ is observed well above the sharp dispersive modes of the pinwheel VBS state.  Thus, while pinwheel dimerization and DM interactions in Rb$_2$Cu$_3$SnF$_{12}$ induce resonant modes at low energies, it appears that a threshold in energy exists beyond which a spin flip is no longer a stable quasi-particle and the underlying quantum kagome nature of the material is apparent.  Whether this scattering is best interpreted as resulting from two-magnon processes or magnon fractionalization will require a more detailed comparison between theories incorporating such features\cite{mourigal} and higher quality scattering data.

\section{Summary}\label{sec4}
High intensity high resolution pulsed neutron scattering unveils new features of the magnetic excitations in the pinwheel VBS state of the distorted kagome lattice antiferromagnet Rb$_2$Cu$_3$SnF$_{12}$.  We observe a weakly dispersive mode around 5~meV and ghost modes, both of which are attributed to the enlarged unit cell caused by the structural transition at $T=215$~K.  Excitations below 8~meV appear to be well described by the dimer series expansion for the enlarged unit cell.  The polarization analysis of the dominant modes is consistent with a splitting of the triplet into a $S_{\mathrm{tot},z}=0$ singlet and a $S_{\mathrm{tot},z}=\pm1$ doublet due to DM interactions.  Between 8~meV and 10~meV, we observe continuum scattering, which is reminiscent of the fractionalized excitations recently observed in herbertsmithite.

\acknowledgments{
The work  was supported in part by the Thailand Research Fund under grant no.\ MRG55800, a Grant-in-Aid for Scientific Research from JPS (nos.\ 23244072, 24740223, and 23540395) and a Global COE Program funded by MEXT Japan. Work at the Institute for Quantum Matter was supported by the US Department of Energy, Office of Basic Energy Sciences, Division of Materials Sciences and Engineering under Award DE-FG02-08ER46544. The research at Oak Ridge National Laboratory's Spallation Neutron Source was sponsored by the Scientific User Facilities Division, Office of Basic Energy Sciences, U.S. Department of Energy.}

\bibliography{RCSF-pinwheel}
\bibliographystyle{apsrev}

\end{document}